\documentclass[%
 reprint,
 amsmath,amssymb,
 aps,prx,
]{revtex4-2}

\usepackage{graphicx}
\usepackage{dcolumn}
\usepackage{bm}

\usepackage{physics}
\usepackage{xcolor}

\begin{document}

\preprint{APS/123-QED}

\title{\textit{Ab initio} calculations of the high-order harmonic enhancement in small noble gas clusters}

\author{Aleksander P. Woźniak}
 \email{ap.wozniak@uw.edu.pl}
\author{Robert Moszyński}%

\affiliation{%
 Faculty of Chemistry, University of Warsaw\\
 Pasteura 1, 02-093 Warsaw, Poland
}%

\date{\today}

\begin{abstract}
We report calculations of the high-order harmonic spectra of few-atomic clusters of two noble gases, helium and krypton, subjected to laser pulses of intensities above $10^{13}$ W/cm\textsuperscript{2}.
We employ a fully \textit{ab intio} framework, namely the real-time time-dependent configuration interaction singles coupled to a discrete Gaussian basis set.
Our model reproduces the observed higher than quadratic dependence between the high-order harmonic yield and the number of atoms, and the obtained value of the enhancement rate is in quantitative agreement with experimental results.
We suggest a mechanism responsible for the high-order harmonic enhancement, according to which, after the tunneling ionization, the residual ion polarizes the electron densities of surrounding atoms in the cluster, which leads to increased electron-electron repulsion during the recollision.
The proposed mechanism is not only fully consistent with both our calculations and the experimental data, but can also explain the limitations of previous theoretical studies on high-order harmonic generation in clusters.
\end{abstract}

\maketitle


\section{Introduction}

High-order harmonic generation (HHG) is a strongly nonlinear optical process in which a gaseous, liquid or solid medium exposed to an intense laser field produces ultrashort pulses of high-frequency radiation in the extreme ultraviolet (XUV) to soft X-ray range.
Thanks to their exceptional spatiotemporal coherence and unprecedented timescales from femto- to attoseconds, these pulses are an invaluable tool in attophysics, as they can be used for investigating the electronic dynamics in atomic or molecular systems \cite{krausz2009,agostini2004}.
A variety of possible applications of high-order harmonic radiation includes, but is not limited to molecular orbital tomography \cite{salieres2012}, determining molecular structures \cite{torres2010}, obtaining vibrational signatures \cite{li2008}, and studying quantum coherence phenomena \cite{marciniak2019,monfared2022}.

Historically, atomic noble gases were the first medium commonly used for HHG due to their high ionization potentials and thus favorable harmonic conversion efficiency \cite{mcpherson1987,ferray1988,sarukura1991,crane1992}.
The mechanism of HHG in single atoms is nowadays well understood, as it has been explained by the well-known three-step model (3SM) of Corkum \cite{krause1992,schafer1993,corkum1993} and Lewenstein \cite{lewenstein1994}.
According to the 3SM a single HHG event initiates with a laser-induced tunneling ionization of the atomic target.
The resulting free electron is driven away by the laser field and then reaccelerated towards the residual ion when the field switches its sign.
In the final step the electron and the ion recombine, which is accompanied by the emission of radiation of frequency equal to some multiple of the driving frequency.
From this model the characteristic shape of the HHG spectrum can also be deduced, consisting of three distinctive regions: (i) the perturbative region characterized by the rapid decrease of intensity of the first few harmonics, (ii) the plateau with nearly constant harmonic intensity and (iii) the cutoff region with an abrupt drop of the harmonic intensity at the cutoff energy $E_{cut} = I_p + 3.17U_p$, where $I_p$ is the ionization potential of the target and $U_p$ is the ponderomotive energy of the driving field.
The predictions of the 3SM were later generalized to polyatomic systems \cite{le2008,suarez2016} and confirmed experimentally in simple molecules \cite{lynga1996}.
However, the common factor of all gaseous media is a relatively low intensity of harmonic radiation caused by their low density \cite{constant1999}.

Another promising candidates for an efficient harmonic generation are Van der Waals clusters of noble gases characterized by high local atomic density comparable to that of solids and liquids that results in very efficient absorption of laser radiation \cite{ditmire1997}.
Noble gas clusters can be obtained under certain pressure and temperature conditions and may exist in a wide variety of sizes, ranging from a few to a few million atoms.
Experimental observation of HHG from clusters was reported first in argon \cite{donnelly1996}, and later also in krypton \cite{ruf2013} and xenon \cite{tisch1997} jets.
Most of the authors observe a significantly increased harmonic yield of clusters when compared to atoms \cite{donnelly1996,tisch1997,vozzi2005,pai2006,aladi2014,park2014,tao2017}, while some of them also report a stronger dependence of the cluster harmonic intensity on the driving field strength \cite{donnelly1996,aladi2014}, suggesting a mechanism of HHG slightly different from the atomic 3SM.
In order to explain this enhancement, several alternative recollision scenarios were proposed that would serve as extensions to the 3SM.
In the ``atom-to-itself" scenario, each atom within the cluster acts as an independent emitter and the ionized electron always returns to its parent ion. In the electron-hole picture, the electron hole, created when an electron is ejected from one of the atoms, remains localized on the residual ion throughout the electron's presence in the continuum.
In the ``atom-to-neighbor" pathway, the electron can also recombine with ions neighboring the parent ion. These adjacent ions can be generated when the electron hole moves from the residual ion to one of the neighboring atoms while the electron is in the continuum, or as a result of these atoms undergoing ionization themselves \cite{veniard2001}.
Finally, in the ``cluster-to-itself" scenario, the ground state electronic wavefunction of the cluster is partially or wholly delocalized over its constituent atoms. Therefore, when the electron is ionized, the resulting electron hole also is spread over the entire cluster volume and the recombination occurs not at particular atomic centers but within the cluster as a whole \cite{ruf2013}.

Aside from increased conversion efficiency, some of the earlier works also report an extension of the cutoff energy in cluster HHG \cite{vozzi2005,aladi2014} that can only be explained with the two latter mechanisms.
According to the three-step model, the cutoff position can be inferred from the lengths of the semiclassical trajectories traversed by the ionized electron.
In an idealized scenario of the ``atom-to-itself” recombination, a single HHG event is confined to only one atom within the entire cluster, where both tunnel ionization and recombination occur, and the presence of the neighboring atoms has no influence on the process.
Therefore, the electron's trajectories and the cutoff position are the same as in the case of an isolated atom.
On the other hand, in the ``neighbor-to-itself” pathway the electron’s trajectory may be extended or shortened by the distance between the atom from which the ionization occurs and the atom at which the recombination takes place.
Similarly, in the ``cluster-to-itself” scenario the electron may be detached from and recombine with different parts of the cluster, which also alters its trajectory.
Therefore, in both of these recombination pathways additional harmonics should be observed beyond the 3SM cutoff of a single atom.
However, more recent studies do not confirm such a cutoff extension \cite{park2014,tao2017}, suggesting that regardless of the cluster size the predominant recombination scenario is ``atom-to-itself" \cite{bodi2019}, while the enhancement of the harmonic intensity may rather be attributed to clusters' increased polarizability and lowered ionization potential \cite{park2014,tao2017}.
Park \textit{et al.} \cite{park2014} observed that the harmonic yield increase is most pronounced in the smallest clusters containing up to a few hundred atoms for which the enhancement rate grows rapidly with size, but clusters beyond some optimal size quickly lose this feature.
This was also confirmed by Tao \textit{et al.} \cite{tao2017}, who suggested that while in small clusters the optical response can be generated in the whole cluster volume, in large clusters only atoms on the surface contribute to HHG due to field screening and reabsorption of the radiation by the cluster core.
Despite these findings, some of the questions still remain open, such as how exactly the interatomic interactions lead to the enhancement of total optical response and how the transition between atomic HHG and cluster HHG occurs.
Addressing the latter issue requires further studies on smallest clusters (of sizes below 100 atoms) that are, alas, difficult to obtain with currently used experimental techniques.

The topic of HHG enhancement in noble gas clusters has also been investigated using various theoretical methods, usually relying on the single-active-electron (SAE) approximation or reduced dimensionality models \cite{hu1997,numico2000,veniard2001,dealdana2001,zaretsky2010,xia2012,park2014,feng2015}.
However, the results of these simulations are sometimes ambiguous and may heavily depend on the assumptions made during the construction of a particular model, such as the effective potential of the cluster atoms and the shape of the initial electronic state.
For example, de Aldana \textit{et al.} \cite{dealdana2001} demonstrated that an extension of the cutoff energy up to $I_p + 8U_p$ can be achieved if the initial electronic wavefunction is distributed over all atoms within a cluster, but harmonics in this energy range were never observed experimentally.
A similar cutoff position was predicted by Hu \textit{et al.} \cite{hu1997}, and later by Véniard \textit{et al.} \cite{veniard2001} and Zaretsky \textit{et al.} \cite{zaretsky2010}, who explained it by the ``atom-to-neighbor" recombination.
However, as pointed out by Véniard \textit{et al.} \cite{veniard2001}, this mechanism should be most effective when the distances between adjacent atoms are close to the quiver amplitude of the ionized electron, which is a value far larger than typical interatomic distances in noble gas clusters.
On the other hand, Park \textit{et al.} \cite{park2014}, who assumed only partial delocalization of the active electron, obtained results consistent with the 3SM cutoff prediction that agreed with their experimental observations.

In this paper we report calculations of the high-order harmonic spectra for smallest noble gas clusters containing less than ten atoms. Aware of the limitations of approximate numerical models, we decided to use a fully \textit{ab initio} framework known based in quantum chemistry, namely the real-time time-dependent configuration interaction singles (RT-TDCIS) with the linear combination of atomic orbitals (LCAO-MO) representation of the electronic wavefunction in a discrete Gaussian basis set.
RT-TDCIS has already been successfully applied to predict HHG spectra of various atomic \cite{luppi2013,coccia2016a,coccia2016b,pabst2016,coccia2019,coccia2020,wozniak2021,wozniak2022} and molecular \cite{luppi2012,white2016,labeye2018,bedurke2019,luppi2021,pauletti2021,morassut2022,wozniak2023} species and becomes an attractive alternative to models based on SAE, especially for systems with high complexity and large numbers of nuclei.
What is worth mentioning, this approach was shown to semi-quantitatively reproduce the ratio between HHG intensities of different compounds \cite{bedurke2019}, which renders it potentially suitable for investigating a similar problem of HHG enhancement in noble gas clusters.
For the main model noble gas we choose helium and consider clusters consisting of four, six and eight He atoms.
Although due helium's low condensation parameter \cite{wormer1989} and very high pressures required for its clusterization, HHG from He clusters has not yet been experimentally observed, the existence of the clusters themselves is well documented in literature \cite{whaley1994}.
From the theoretical point of view helium is particularly advantageous as it has only two electrons, which allows for a fully explicit treatment of even the largest of the considered clusters at a reasonable computational cost.
However, in order to further validate that our findings for helium translate well to heavier noble gases, we also perform calculations for the smallest tetraatomic cluster of krypton.

The paper is constructed as follows: In Sec. II we provide an outline of theory. In Sec. III we present and discuss the numerical results. Finally, in Sec. IV we summarize our work.

\section{Methods}

RT-TDCIS provides an approximate solution to the time-dependent Schr\"odinger equation (the atomic units are used throughout the paper),
\begin{equation} \label{tdse}
i \frac{\partial}{\partial t}\Psi(t) = \hat{H}(t)\Psi(t),
\end{equation}
by expanding the time-dependent $n$-electronic wavefunction $\Psi(t)$ in the basis of the time-indepenent CIS states $\Psi_k$,
\begin{equation} \label{rttdcis}
\Psi(t) = \sum_k C_k(t) \Psi_k.
\end{equation}
The CIS ground state $\Psi_0 = \Phi_\mathrm{HF}$, the Hartree-Fock ground (reference) determinant. Since we describe the time-evolution of closed-shell systems without any spin-dependent perturbation, we only need to consider singlet excited states $\Psi_{k>0}$, which are linear combinations of all singly-excited singlet-state configurations $\Phi_i^a$,
\begin{equation} \label{cisstate}
\Psi_{k>0} = \sum_{i}^{n_{occ}} \sum_{a}^{n_{vir}} c_{i,k}^a \Phi_i^a.
\end{equation}
where the summation is over $n_{occ}$ occupied and $n_{vir}$ virtual orbitals.
The excited states are obtained by solving the eigenequation of the ground state CIS Hamiltonian, $\hat{H}_0 c_k = E_k c_k$, where $E_k$ is the energy of $k$-th electronic state.
According to the Brillouin's theorem, there is no mixing between the Hartree-Fock determinant and the excited configurations, therefore $E_0 = E_\mathrm{HF}$.

In this work, the time-dependent Hamiltonian $\hat{H}(t)$ is represented in the length gauge, as a sum of the ground state Hamiltonian and the electric field operator in the dipole approximation,
\begin{equation}
\hat{H}(t) = \hat{H}_0 - \vec{\hat{\mu}}\vec{\mathcal{E}}(t),
\end{equation}
where $\vec{\hat{\mu}}$ is the total dipole moment operator of the considered system and $\vec{\mathcal{E}}(t)$ is the time-dependent electric field vector representing the external, linearly polarized laser field.
In the calculations an electric field pulse with a sine-squared envelope is used,
\begin{equation} \label{efield}
\vec{\mathcal{E}}(t) = 
\begin{cases}
\vec{\mathcal{E}}_0 \sin(\omega_0 t) \sin^2(\omega_0 t/2 n_c)&\text{if}\; 0 \leq t \leq 2\pi n_c/\omega_0, \\
0& \text{otherwise,}
\end{cases}
\end{equation}
where $\vec{\mathcal{E}}_0$ is the field amplitude vector, $\omega_0$ is the carrier frequency and $n_c$ is the number of optical cycles.

We propagate the electronic wavefunction using the second-order split operator. The $\hat{H}_0$ matrix in the basis of the CIS eigenstates is a diagonal matrix of eigenenergies, $\mathbf{H_0}_{kl} = \delta_{kl}E_k$, while the dipole moment operator has three spatial components, $\vec{\hat{\mu}} = [\hat{\mu}_x, \hat{\mu}_y, \hat{\mu}_z]$. Therefore the final matrix propagation equation in the CIS basis reads
\begin{equation} \label{propagation}
\begin{split}
\mathbf{C}(t+\Delta t) &= e^{i \mathbf{H_0}\Delta t/2} \times \\
& \times \left[ \prod_{j=x,y,z} \mathbf{U}_j^\dag e^{i \mathcal{E}_j(t)\mathbf{d}_j \Delta t} \mathbf{U}_j \right] e^{i \mathbf{H_0}\Delta t/2} \mathbf{C}(t),
\end{split}
\end{equation}
where the unitary matrix $\mathbf{U}_j$ transforms the $j$-th dipole component matrix $\boldsymbol{\mu}_j$ from the CIS basis to the basis in which it is diagonal, $\boldsymbol{\mu}_j = \mathbf{U}_j^\dag \mathbf{d}_j \mathbf{U}_j$.

The molecular orbitals (MOs) used for the construction of the reference determinant and the singly excited configurations are obtained by solving the restricted Hartree-Fock (RHF) equations in a predefined, atom-centered Gaussian basis set. In the calculations of the helium clusters we use the aug-cc-pVQZ Dunning basis set for He which we further supplement with the active range-optimized (ARO) functions introduced by us \cite{wozniak2021} to describe highly excited and continuum states of atoms and molecules. To the basis set of each atom we add a set of three ARO functions per every angular momentum already present in the aug-cc-pVQZ basis set ($l=0,1,2,3$). The ARO functions are obtained according to the procedure described in Ref. \cite{wozniak2021}, by fitting the Gaussian exponents to the set of Slater orbitals with $n$ from 1 to 100 and $\zeta = 1$. The values of the exponents are collected in the Supplemental Material \cite{sm}.

In order to compensate for the incompleteness of the basis set we employ the heuristic lifetime model of Klinkusch \textit{et al.} \cite{klinkusch2009}. To every CIS state beyond the ionization threshold we add an imaginary ionization rate, $E_k \rightarrow E_k - i \Gamma_k /2$. It is calculated as a sum of the semiclassical ionization rates of virtual orbitals, weighted by the contributions of the excited configurations containing these orbitals to a given state,
\begin{equation}
\Gamma_k = \sum_{i}^{occ.} \sum_{a}^{vir.} |c^a_{i,k}|^2 \theta(\epsilon_a) \sqrt{2\epsilon_a}/d.
\end{equation}
Here, $\epsilon_a$ is the energy of $a$-th virtual orbital, $\theta(x)$ is the Heaviside function (ensuring that only orbitals with $\epsilon_a > 0$ are ionizable) and $d$ is a characteristic escape length, after travelling which the electron is considered ``free". After inserting these modified complex energies into Eq. \eqref{propagation}, the CIS states above $I_p$ are assigned finite lifetimes and the norm of the time-dependent wavefunction decreases over time, simulating ionization losses. The ionization potential in the model is approximated from the Koopmans theorem, as $-\epsilon_{\mathrm{HOMO}}$.

After performing the time propagation we calculate the HHG spectrum in the velocity form, as the squared modulus of the Fourier transformed time-resolved dipole velocity $v(t) = \frac{d}{dt}\langle\mu\rangle(t)$, normalized by the total propagation time $T$,
\begin{equation} \label{hhg}
I_{\mathrm{HHG}}(\omega) = \left| \frac{1}{T}\int_{0}^{T}W(t)v(t) e^{i\omega t} dt \right|^2.
\end{equation}
The components of the time-resolved dipole velocity are calculated from
\begin{equation}
v_j(t) = -i \sum_{k,l} c_l^\dag(t) c_k(t) \langle\Psi_l\vert \hat \partial_j \vert\Psi_k\rangle, \quad j \in \{x,y,z\},
\end{equation}
and $W(t)$ is the Hann window function applied to account for the finite simulation time.

To investigate the harmonic enhancement in the cluster HHG we have to perform calculations for both the clusters and the single noble gas atom.
However, it is well known that in the LCAO-MO approach the quality of the electron dynamics description is extremely sensitive to the choice of the basis set.
Doing the computations for the atom in a single-centered basis set may be thus suboptimal, because the lack of mixing between functions at different centers (possible in a cluster basis set) will lead to a much worse representation of the time-dependent wavefunction, making the atomic and cluster spectra incomparable.
To avoid this we employ a different strategy, similar to the counterpoise correction used in the single-point calculations of intermolecular properties \cite{boys1970}.
For every considered cluster geometry we obtain the corresponding optical response of the single atom by performing the propagation of the atomic wavefunction in the basis set of this cluster, i.e. we remove all atoms except one and leave the basis functions of the removed atoms in a form of ghost atoms.
In case of an $N$-atomic cluster, we repeat the propagations for all $N$ possible positions of the ``physical" atom, and then average the resulting time-resolved observables to get the atomic dipole velocity.
This approach, dubbed ``ghost-averaging" for short, apart from reducing basis set inconsistencies, has an additional advantage of making the CIS space approximately size-extensive with respect to the number of atoms.
The dimension of the CIS Hamiltonian matrix is $n_{occ} \times n_{vir} + 1$, so for $n_{vir} \gg n_{occ}$ -- which is usually desired for the reasonable approximation of the electronic continuum -- the CIS space of an $N$-atomic cluster tends to $N$ times the CIS space of the single atom in the basis set of the cluster.

Furthermore, to account for different possible orientations of the clusters with respect to the laser field, we perform the calculations at 26 different electric field polarizations along the unit vectors $ \vec{\mathbf{e}}_i = \mathcal{N}[e_x,e_y,e_z]$, where $e_{j=x,y,z} \in \{-1,0,1\}$ (excluding the trivial case of $[0,0,0]$), and $\mathcal{N} = (e_x^2+e_y^2+e_z^2)^{-1/2}$ is the normalization constant. 
The time-resolved dipole velocity vectors (ghost-averaged in case of atomic simulations) $\vec{\mathbf{v}}_i(t)$ from all propagations are then projected onto respective unit polarization vectors and averaged,
\begin{equation}
\bar{v}(t) = \frac{1}{26} \sum_{i}^{26} \vec{\mathbf{v}}_i(t) \cdot \vec{\mathbf{e}}_i
\end{equation}
and the resulting mean dipole velocity $\bar{v}$ is inserted into Eq. \eqref{hhg} to obtain the rotationally averaged HHG spectrum.

In the calculations, we use the Dalton2020.0 software \cite{dalton} to generate the necessary one- and two-electron integrals between the basis functions.
All the remaining parts of the computations: solving the RHF problem, full diagonalization of the CIS matrix and the real-time propagation, are performed using a home-made suite of programs.
When performing calculations, we found out that the large numbers of diffuse functions in the applied basis sets, unnatural for systems comprised of noble gas atoms, proved challenging for the convergence of the standard self-consistent field (SCF) algorithm.
For that reason we employed a specifically tailored RHF procedure, in which we first solve the RHF equations in the pure Dunning basis set, and then use the obtained density matrix as a starting guess for the solution of the RHF problem in the full (Dunning+ARO) basis set.

\section{Results and discussion}

\subsection{Helium clusters}

\begin{figure}
\centering
\includegraphics[width=0.99\linewidth]{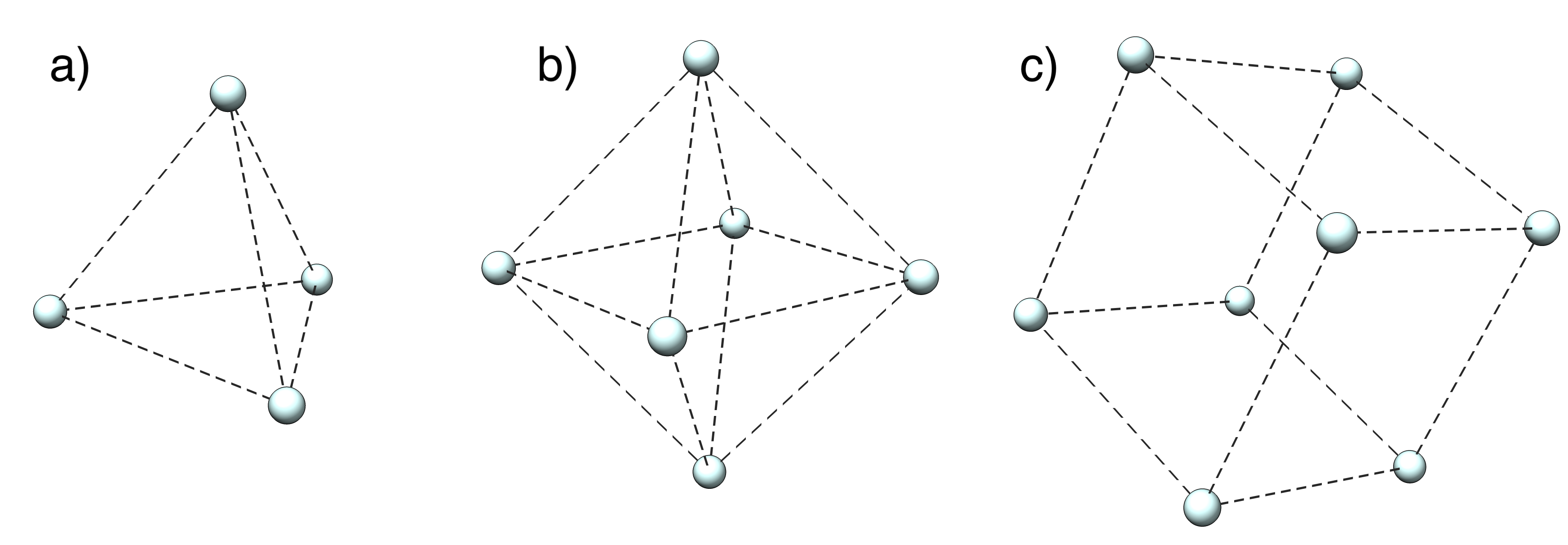}
\caption{Geometries of the studied helium clusters: (a) He\textsubscript{4}, (b) He\textsubscript{6}, (c) He\textsubscript{8}; the dashed lines mark the distances between closest He atoms, set to 3.6 \AA}
\label{fig:clusters}
\end{figure}

In this subsection we present the results of the calculations for three helium clusters consisting of four, six and eight atoms, arranged in tetrahedral, octahedral and cubic manner, respectively (Fig. \ref{fig:clusters}).
The distance between closest He atoms is set to 3.60 \AA, which is the experimental value of the average He-He interatomic distance in liquid helium nanodroplets \cite{peterka2007}.

The electric field pulse consists of 20 optical cycles, with carrier frequency of 1.55 eV (corresponding to a wavelength of 800 nm), so that the its total duration is approx. 2206 a.u. (53.4 fs).
The cycle-averaged laser intensity $I_0$, related to the field amplitude via $I_0 = \mathcal{E}_0^2/2$, is set to $2\times 10^{14}$ W/cm\textsuperscript{2}.
We propagate the wavefunction with the timestep $\Delta t$ = 0.01 a.u. ($\approx$ 0.24 as).
The finite lifetime model parameter $d$ is set to 50 bohr ($\approx$ 26.5 \AA), a value that exceeds the electron quiver amplitude under considered laser conditions.
For every considered cluster geometry we also perform a corresponding calculation for a single He atom, using the ghost-averaging scheme described in the previous section.
After the necessary removal of linear dependencies in the basis sets at the SCF step, we arrive at RT-TDCIS propagation equations of orders 1489, 3271 and 5313 for He\textsubscript{4}, He\textsubscript{6} and He\textsubscript{8}, respectively, and 376, 551 and 672 for the single He atom in the basis of He\textsubscript{4}, He\textsubscript{6} and He\textsubscript{8}, respectively.

\begin{figure*}
\centering
\includegraphics[width=0.99\linewidth]{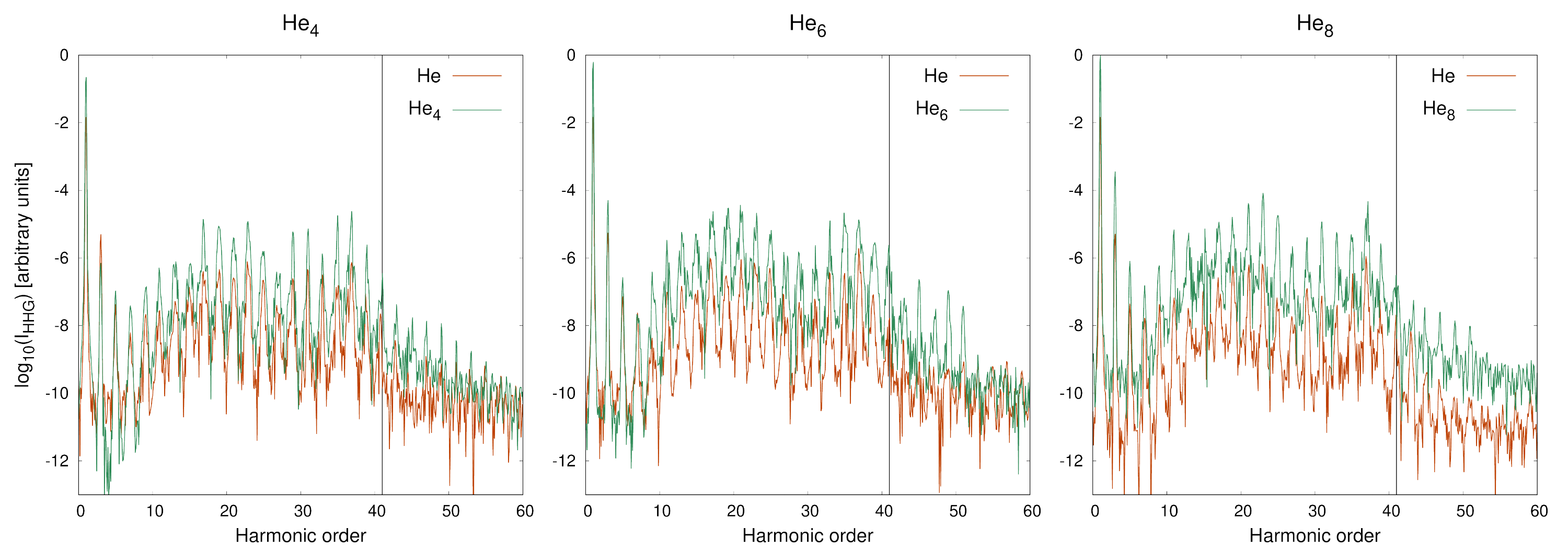}
\caption{Computed HHG spectra of atomic helium and helium clusters. The black vertical lines denote the 3SM cutoff position}
\label{fig:he_spectra}
\end{figure*}

\begin{figure}
\centering
\includegraphics[width=0.99\linewidth]{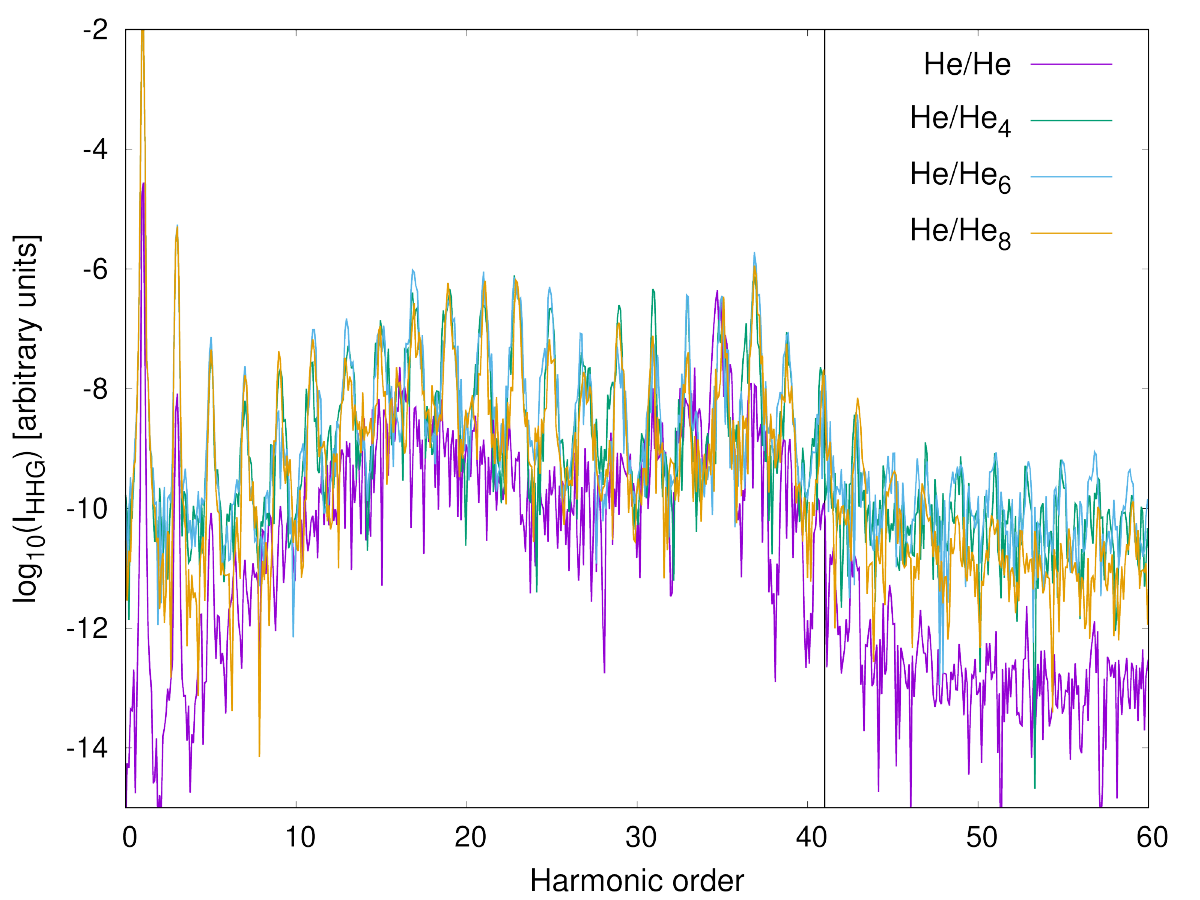}
\caption{HHG spectra of atomic helium computed in the single-centered aug-cc-pVQZ+ARO basis and in basis sets of three considered clusters (denoted as [system]/[basis set])}
\label{fig:he_basis}
\end{figure}

The calculated HHG spectra are shown on Fig. \ref{fig:he_spectra}.
At this point we present them in an unmodified form, i.e. without dividing the HHG signal from clusters by any power of the number of constituent atoms.
Such a representation best mimics the data acquired in experiments, where the HHG signal is collected at different atomic densities, but the volume of interaction between the laser beam and the target remains constant, resulting in different amounts of atoms in the sample.
As a preliminary remark, it can be noticed that the atomic spectra obtained using different cluster basis sets vary slightly from one another, which is an unavoidable drawback of the ghost-averaging approach.
To check whether all three basis sets provide a comparable quality in describing the electronic dynamics, on Fig. \ref{fig:he_basis} we compare the atomic spectra from Fig. \ref{fig:he_spectra}, along with a spectrum of the He atom calculated using a single-centered aug-cc-pVQZ+ARO basis set.
One can immediately notice a substantial difference in the description of the atomic HHG when transitioning from the single-centered to the four-centered basis set, with the former being unable to reproduce the majority of the harmonic peaks.
On the other hand, the spectra obtained using three multicentered basis sets are extremely similar to each other, especially in the plateau region.
The differences in the peak intensities in the spectra obtained using different cluster basis sets are, in fact, negligible compared to the differences in peak intensities between every atomic spectrum and its corresponding cluster spectrum on Fig. \ref{fig:he_spectra}.
This indicates that expanding the basis set from the single-centered one to the four-centered one is sufficient to achieve resolution convergence, and further addition of ghost atoms has minimal effect on the description of HHG.
However, we would like to emphasize that while ensuring the proper accuracy of the basis sets is an important factor in validating the correctness of our model, from the perspective of this work, it is far more crucial for each cluster spectrum and its corresponding atomic spectrum to be computed using bases of identical resolution, which, in our case, is ensured by the use of ghost atoms.

The first important observation regarding the spectra on Fig. \ref{fig:he_spectra} is that the cutoff positions of both atomic and cluster spectra are in good agreement with the prediction of the three-step model.
In the atomic HHG the cutoff is well localized, as the intensities of the peaks beyond the 41st harmonic (3SM cutoff position) quickly drop to the level of numerical noise.
In the clusters the decrease of the harmonic intensity is marginally slower, which results in the cutoff region being more spread over a few consecutive harmonic orders.
This is best seen in the spectra of He\textsubscript{6} and He\textsubscript{8}, where additional peaks can be seen up to 50th harmonic order, although their intensities are a few orders of magnitude lower compared to the ones in the plateau region.
Overall, the range of this cutoff extension is nowhere near the value of $I_p + 8U_p$ reported by earlier theoretical works, and remains consistent with the experimental data for heavier noble gases, indicating that HHG in small clusters is still dominated by the three-step mechanism with ``atom-to-itself" recombination.

\begin{figure}
\centering
\includegraphics[width=0.99\linewidth]{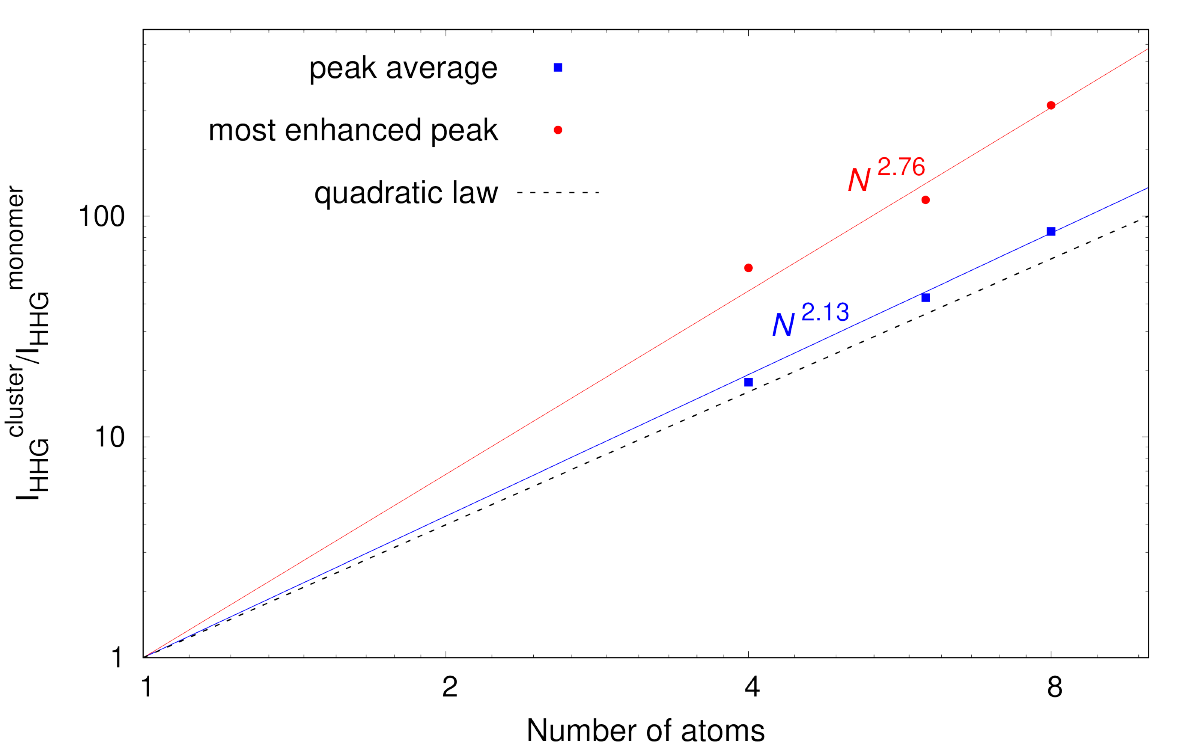}
\caption{Dependence of the HHG intensity of helium clusters on the number of atoms, calculated from the average plateau peak intensity (blue dots) and from the intensity of the plateau harmonic that exhibits strongest HHG enhancement (red dots). The solid lines are the power functions $N^k$ fitted to each data set, with their respective equations shown next to them}
\label{fig:he_enhancement}
\end{figure}

In order to determine and quantify the presence of the HHG enhancement in the obtained cluster spectra, we first have to recall that the HHG signal depends quadratically on the optical response of the system, which in our case is chosen to be the dipole velocity (Eq. \eqref{hhg}).
Therefore if there is no enhancement, the HHG signal of a system of $N$ non-interacting atoms, such as an ensemble of noble gas atoms in a gas phase, should scale as $N^2$ times the HHG signal of a single atom, a property that is sometimes referred to as the ``quadratic law" \cite{veniard2001,park2014}.
This relies on highly idealized assumptions that the optical responses of all atoms add up perfectly coherently, and that there is no reabsorption of the emitted radiation.
While both of these conditions are met in our model, they are obviously not satisfied in real physical conditions, so the experimentally measured dependence is often lower that $N^2$ \cite{aladi2014,park2014}.
The steeper increase of the harmonic yield with atomic density in clusters compared to atoms is usually observed in experiments as an abrupt change in the slope of the log-linear relationship between the backing pressure and the HHG signal, which accompanies the cluster formation.
This behavior is commonly interpreted as a transition from the quadratic law to a higher than quadratic dependence between the number of atoms and the HHG intensity, $N^k$ with $k>2$ \cite{tisch1997,park2014,tao2017}.
In order to allow a comparison with experimental findings, we apply the same model to the obtained results.
To verify the presence the harmonic enhancement, we calculate the ratios between the average peak intensities in the plateau region (harmonics 9th to 41st) of the cluster HHG spectra and of the corresponding atomic HHG spectra.
We then fit a power function $N^k$ to the generated set of points, and obtain the value of $k=2.13$ (Fig. \ref{fig:he_enhancement}).
This confirms a relatively small, yet significant -- given that all the points lay above the quadratic line curve -- HHG enhancement in the helium clusters.
However, in the experimental works the dependence between the HHG signal and $N$ is not usually measured for all plateau harmonics, but only for one or a few of them that are characterized by highest overall yield or exhibit the strongest level of enhancement \cite{donnelly1996,tisch1997,pai2006,aladi2014,park2014,tao2017,bodi2019}.
Therefore, from each spectrum we also pick the harmonic peak with the highest cluster-to-atom intensity ratio, and perform a second power fit, this time obtaining a larger value of $k=2.76$.

At this point it is necessary to discuss the limitations of the applied model, with a particular focus on the lack of higher excitations within the RT-TDCIS approach.
From the perspective of the laser-driven electron dynamics, excitations higher than single are required for the description of two distinct classes of phenomena: the dynamic correlation between the electrons in the residual ion and the ejected electron, and the processes which involve the detachment of more than one electron from the system, such as non-sequential double ionization.
The former subject has been extensively studied theoretically by us and by other authors \cite{nguyen2006,castro2015,artemyev2017,sato2018,neufeld2020,reiff2020,saalfrank2020,pauletti2021,wozniak2022,wozniak2023}, with an overall conclusion that although the inclusion of dynamical correlation has some subtle effects on the HHG process, the majority of the features of HHG can be recovered using a non-correlated approach.
The issue of multiple ionization, on the other hand, feels more relevant to this study, as RT-TDCIS is unable to describe a scenario in which two or more atoms within the cluster become ionized at the same time.
However, our results clearly suggest that HHG in noble gas clusters can be described using a model with only one electron per whole cluster undergoing tunelling ionization.
This indicates that at some point during the formation of small clusters there is a transition from the mechanism in which all atoms act as independent harmonic emitters to the molecular-like mechanism in which the cluster becomes only singly ionized.
Otherwise, an $N$-tuply excited wavefunction would be required to reproduce not only the HHG enhancement, but also the quadratic law in an $N$-atomic cluster.

In order to get more insight into the possible mechanism of the HHG enhancement, we analyzed the clusters' ionization potentials (estimated by $-\epsilon_{\mathrm{HOMO}}$), as well as static dipole polarizabilities and dynamic dipole polarizabilities at 800 nm, as these properties are most frequently cited as associated with the harmonic yield increase \cite{park2014,tao2017}.
All the properties were calculated at the RHF level of theory, using the same basis sets as in the real-time propagations (Table \ref{tab:properties}).
It can be seen that both the static and dynamic polarizability of every investigated He cluster is practically equal to the sum of polarizabilities of isolated He atoms.
Moreover, the examination of the ground state wavefunctions has shown that the molecular orbitals of He clusters are to a great extent linear combinations of molecular orbitals of isolated He atoms.
This is further evidenced by extremely small differences between the energies of highest and lowest occupied MOs in all considered clusters, also given in Table \ref{tab:properties}.
The observed lack of mixing between the MOs of constituent atoms should not be surprising, considering that we perform calculations for Van der Waals bonded systems using a method that neglects dynamical correlation, and hence cannot describe dispersion interactions.
However, the fact that even when using an uncorrelated approach, that may underestimate orbital overlapping between different atoms, we can successfully reproduce the HHG enhancement, suggests that this phenomenon should not be attributed to the ground-state correlation or its delocalization.
These observations indicate the absence of any excess atomic polarizability within the clusters that could be responsible for the increase of the harmonic yield, as well as provide additional evidence against the ``cluster-to-itself" recombination scenario, favoring the ``atom-to-itself" one.
More precisely, the obtained results allow us to postulate a specific variant of the ``atom-to-itself" recombination, in which the ejected electronic wavepacket is a coherent superposition of states, each of which describes the detachment of an electron from a different atom within the cluster.
It should be noted that in the considered clusters each atom contributes equally to every cluster MO due to high symmetry.
If symmetry is lowered, it may happen that the tunelling ionization will occur more likely in some atoms than others, and will have a more ``localized" character.

\begin{table*}
\caption{\label{tab:properties}
Ground state properties of helium atom and clusters calculated at the RHF level of theory using the aug-cc-pVQZ+ARO basis sets: ionization potential (estimated from the Koopmans theorem), difference between highest and lowest occupied MO energy, static dipole polarizability per atom, and dynamic dipole polarizability per atom calculated at 800 nm ($\omega$ = 0.05695 a.u.) }\begin{ruledtabular}
\begin{tabular}{ccccc}
 &$I_p$ (eV) &$\Delta_{{\mathrm{HOMO}}-{\mathrm{LOMO}}}$ (eV)& $\alpha(0.0)/N$ (a.u.\textsuperscript{3})\footnotemark[2] & $\alpha(0.05695)/N$ (a.u.\textsuperscript{3})\footnotemark[2]\\
\hline
He\footnotemark[1] & 24.978 & -- & 1.321 & 1.326 \\
He\textsubscript{4} & 24.965 & 0.052 & 1.321 & 1.326 \\
He\textsubscript{6} & 24.952 & 0.078 & 1.321 & 1.325 \\
He\textsubscript{8} & 24.939 & 0.078 & 1.321 & 1.325 \\

\end{tabular}
\end{ruledtabular}
\footnotetext[1]{The presented values for single He atom are calculated in the basis set of the He\textsubscript{8} cluster since it is the largest one used, however additional calculations in the basis sets of He\textsubscript{4} and He\textsubscript{6} gave the same results up to the precision used.}
\footnotetext[2]{$\alpha(0.0)$ and $\alpha(0.05695)$ are one of the diagonal elements of the respective static and dynamic polarizability tensor (for each considered system $\alpha_{xx} = \alpha_{yy} = \alpha_{zz}$ for symmetry reasons).}
\end{table*}

\begin{figure*}[h]
\centering
\includegraphics[width=0.99\linewidth]{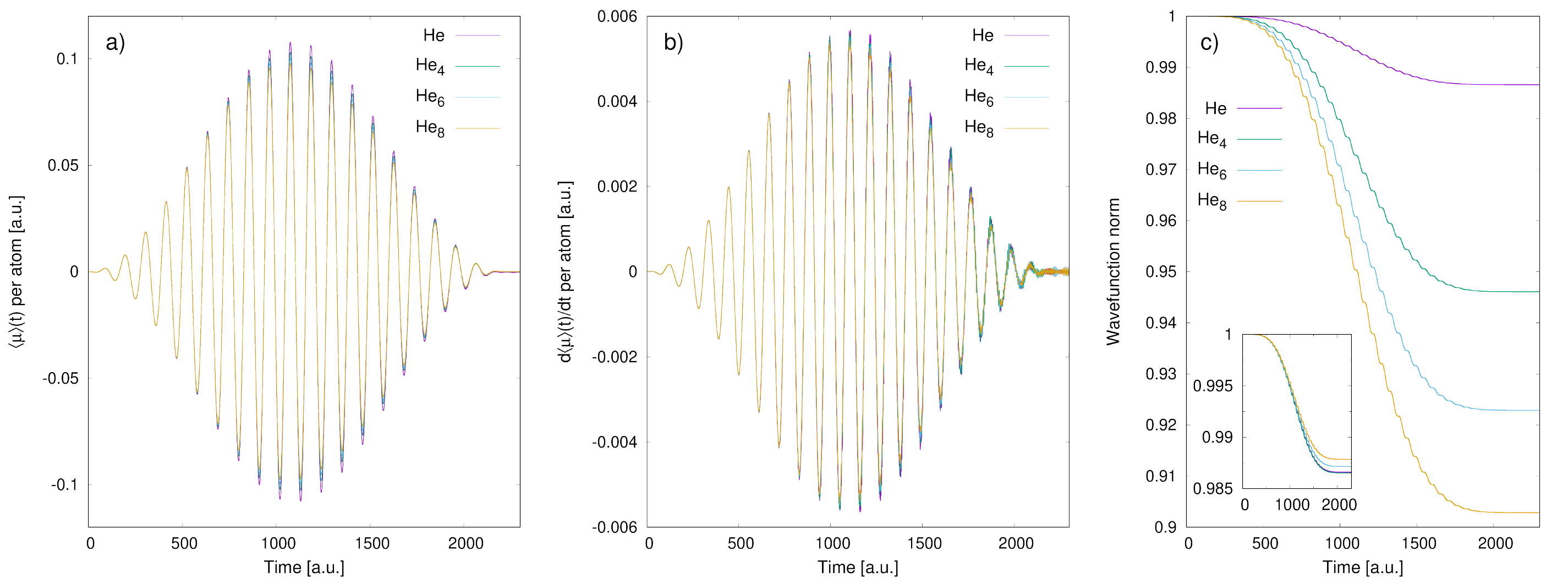}
\caption{a) Time-resolved dipole moment per atom, b) time-resolved dipole velocity per atom, and c) time-resolved total wavefunction norm from the RT-TDCIS calculations of the helium atom and clusters. The inset in c) shows the time-resolved wavefunction norm per atom}
\label{fig:he_obser}
\end{figure*}

In Table \ref{tab:properties}, one can also notice a minor reduction of the first ionization energy of helium as the cluster size increases.
This trend, although systematic, is nonetheless extremely small, with the difference between $I_p$ of a single He atom and the He\textsubscript{8} cluster of less then 0.04 eV.
Such an outcome is consistent with the obtained HHG spectra, because if both the atomic and cluster HHG occurs via the three-step mechanism with single ionization, any significant change in $I_p$ would translate to a shift in the cutoff position according to the $I_p + 3.17U_p$ formula.
When analyzing the time-resolved observables, we observe that the amplitudes of the average atomic contribution to the dipole moment and dipole velocity tend to slightly decrease with increasing $N$ as well (Fig. \ref{fig:he_obser}a nd b).
This effect could be partially attributed to the lack of multiple excitations in the CIS excited states.
However, since it becomes most prominent after the external field reaches its maximum value, it is more likely to be caused by the decrease in the norm of the time-dependent wavefunction (Fig. \ref{fig:he_obser}c), which follows a similar tendency.
Although the increase in the total ionization probability with the cluster size may intuitively appear to be connected to the lowering of the ionization potential, we argue that it is rather a consequence of the discrete and finite nature of the CIS energy eigenspectrum and a somewhat crude description of the ionization process by the finite lifetime model.
The size of the CIS space, and hence also the amount of available excited states, scales more or less linearly with the number of atoms.
Thus, larger clusters are characterized by a greater number of states with non-zero ionization rates.
The larger total polarizabilities of clusters compared to individual atoms allow more of these states to become populated in the laser field, which leads to higher ionization losses.
This becomes evident when comparing the differences in ionization potentials and total ionization probabilities.
The difference in $I_p$ between He and He\textsubscript{4} is exactly the same as between He\textsubscript{4} and He\textsubscript{6}, so if the ionization potential was responsible for the increased total ionization probability of clusters, we would expect to observe a similar pattern in the wavefunction norm profiles.
Meanwhile, the difference in the total ionization probability between He and He\textsubscript{4} is approximately two times larger than between He\textsubscript{4} and He\textsubscript{6}, suggesting that this quantity grows linearly with $N$.
On the inset in Fig. \ref{fig:he_obser}c we can notice that the ionization probabilities per atom are equal for He and He\textsubscript{4}, and actually slightly lowered for He\textsubscript{6} and He\textsubscript{8}, probably reflecting the fact that the CIS space size is not exactly proportional to $N$, and that the distribution of the excited states may vary between systems.
However, we would like to note that the time-resolved norm of the RT-TDCIS wavefunction is not an additive property and it is the total, rather than atomic, ionization probability that should be considered for its potential effect on the HHG enhancement.
Also, the finite lifetime model, just like all other absorbing methods, is meant to simulate these ionization regimes in which the electronic wavepackets are irreversibly removed from the systems and thus do not contribute to HHG, i.e. above threshold ionization and barrier suppression ionization.
Consequently, higher total ionization probability should suppress the HHG intensity, just as it decreases the amplitudes of the atomic optical response.
The fact that we observe the HHG enhancement in our results indicates that this effect is probably relatively weak (in all systems the total ionization loss is less than 10\%) or affects mostly the perturbative region of the spectra.

In the course of the above analysis, we have established that both the ionization potential of the clusters and their polarizability are unlikely to have a significant influence on the increase in the HHG intensity beyond the quadratic law.
Therefore, it is reasonable to consider investigating an alternative mechanism that could be responsible for the boost in the harmonic yield.
It should be also kept in mind that in the light of our results, a single cluster should be viewed as a unitary HHG emitter undergoing single tunneling ionization, and not as an ensemble of atoms responding to the external field.
Here, we present a putative mechanism of the HHG enhancement.
It is conceptually similar to the one proposed by Gordon \textit{et al.} in Ref. \cite{gordon2006} to explain why heavier noble gases emit stronger harmonic radiation then lighter ones even when subjected to laser conditions that ensure similar ionization rates.
Using both analytical and numerical arguments, they demonstrated that in a many-electron system all electrons, both ionized and bound, contribute to the emission of harmonic radiation via so called ``polarizational recombination".
When an electron ejected from an atom with atomic number $Z$ returns to the residual ion, it deaccelerates due to the repulsive force exerted by $Z-1$ bound electrons.
However, the exact same force is exerted by the recolliding electron on the bound electrons, causing them to accelerate as well.
Since every accelerating charge is a source of electromagnetic radiation, all $Z$ electrons become HHG emitters in this picture, even though only one of them actually undergoes tunneling ionization.
Naturally, this effect can only be described at the level of a theory that allows for the motion of all electrons, thus it is possible in RT-TDCIS as well as in any other many-body theory but not in SAE-based models.
Since the movement of $Z$ electrons in a field of a screened nuclear charge of $1$ is equivalent to the movement of one electron in field of an unscreened nuclear charge of $Z$, it becomes evident how the atomic number (or the total number of electrons) is another factor, alongside the polarizability and the ionization potential, that influences the HHG intensity. 
Our mechanism relies on a similar premise.
The additivity of the atomic polarizabilities within the cluster allows for a larger portion of the wavefunction to become ionized, making its contribution to the total dipole moment and velocity proportional to $N$.
In our model, it also results in an approximately linear increase of the total ionization probability with the number of atoms.
However, the increased polarizability also affects the part of the electronic density that remains in the cluster cation after the ionization, making it more susceptible to deformation by the electric field.
Therefore, in a semiclassical picture, an atom within a cluster that became ionized may polarize the neighboring atoms and ``borrow" some of their electron density, increasing the shielding of its own nucleus and dissipating the electron hole over a larger volume.
As a result, the returning electron during the recollision repels not just $Z-1$ electrons as in the case of the isolated atom, but slightly more, between $Z-1$ and $Z$.
Thus, in compliance with the ``polarizational recombination" model, slightly more than $Z$ electrons are involved in the HHG process, which leads to the emission of additional radiation.
The polarization of the neighboring atoms by the residual ion is expected to be most efficient within the first few atomic shells surrounding the ionized atom, but quickly vanish at larger distances.
Therefore, in smallest few-atomic clusters the shielding of the residual ion should be proportional to $N$, as all atoms may contribute to the charge dissipation, but when the cluster size increases, this relationship should gradually weaken, at some point reaching a constant value.
Such a dependence is perfectly consistent with the observed vanishing of the HHG enhancement in large clusters.
Of course in this work we only propose the mechanism responsible for the increase of the harmonic signal, while in actual large clusters there are some additional effects, such as the screening of the external field by the cluster, that further affect the observed HHG intensity.

Our mechanism also explains why in the numerical models of clusters based on SAE the HHG enhancement can be reproduced only by introducing the delocalization of the initial electronic state, even though multiple heuristic arguments, referring e.g. to the weakness of London dispersion forces between noble gas atoms, find such delocalization improbable.
In these models atoms other than the one undergoing ionization are usually replaced with some fixed potentials, such as the soft-Coulomb potential \cite{hu1997,numico2000,dealdana2001,xia2012,park2014,feng2015}.
Therefore, the shielding of the residual ion is impossible to describe if the initial state is confined to only one atom.
However, if the initial state is distributed across multiple atoms, the polarizability of the whole system is artificially increased, so the shielding and the HHG enhancement may occur, but due to an unphysical shape of the electronic wavefunction, this comes with the cost of generating spurious harmonics beyond the 3SM cutoff.
An excellent example is the work of Park \textit{et al.} \cite{park2014}, who achieved a balance between the correct cutoff description and the HHG enhancement by delocalizing the initial wavefunction only over the nearest neighbors of the ionized atom.
Our results show that no initial delocalization is needed (as actually it is not predicted by the \textit{ab initio} solution), if the mobility of electrons in all atoms is explicitly considered.

\subsection{Krypton clusters}

Although our results for the He clusters seem to provide a consistent picture of the HHG enhancement, we also want to check if the mechanism we propose remains valid also for heavier noble gases. 
Since the condensation parameter of noble gases increases with their atomic number \cite{wormer1989}, we chose to perform calculations for krypton, in which the cluster HHG enhancement is among the easiest to observe experimentally.
Due to krypton's much larger number of electrons, we had to limit our calculations only to the smallest four-atomic cluster.
Even for such a small system the dimension of the CIS space including excitations from all occupied orbitals is prohibitively large, so we further reduce the size of the problem by replacing the ten inner core electrons of krypton with a small core relativistic ECP10MDF pseudopotential \cite{peterson2003}.
To describe the outer core and valence electrons we use the aug-cc-pVTZ-PP basis set compatible with this pseudopotential \cite{peterson2003}, which is also supplemented with the same three ARO functions per angular momentum as in the calculations for He clusters.
The orders of the resulting CIS matrix equations are 18721 for Kr\textsubscript{4} and 5188 for a single Kr atom in the basis of Kr\textsubscript{4}.
They are still quite sizeable compared to these of the largest considered He clusters, but computationally feasible.

The four krypton atoms are arranged in a tetrahedral manner, like in the He\textsubscript{4} cluster.
The distance between atoms is set to 4.04 \AA, which is the experimental value of the average Kr-Kr distance in krypton clusters \cite{konotop2023}.
The ionization potential of krypton is much lower than of helium, so we lower the laser intensity to $4\times 10^{13}$ W/cm\textsuperscript{2} and increase the wavelength to 1200 nm to avoid ionization saturation.
In order to maintain the same pulse duration, we reduce the number of optical cycles to 15.
The $d$ parameter in the finite lifetime model is kept at 50 a.u.

\begin{figure}
\centering
\includegraphics[width=0.9\linewidth]{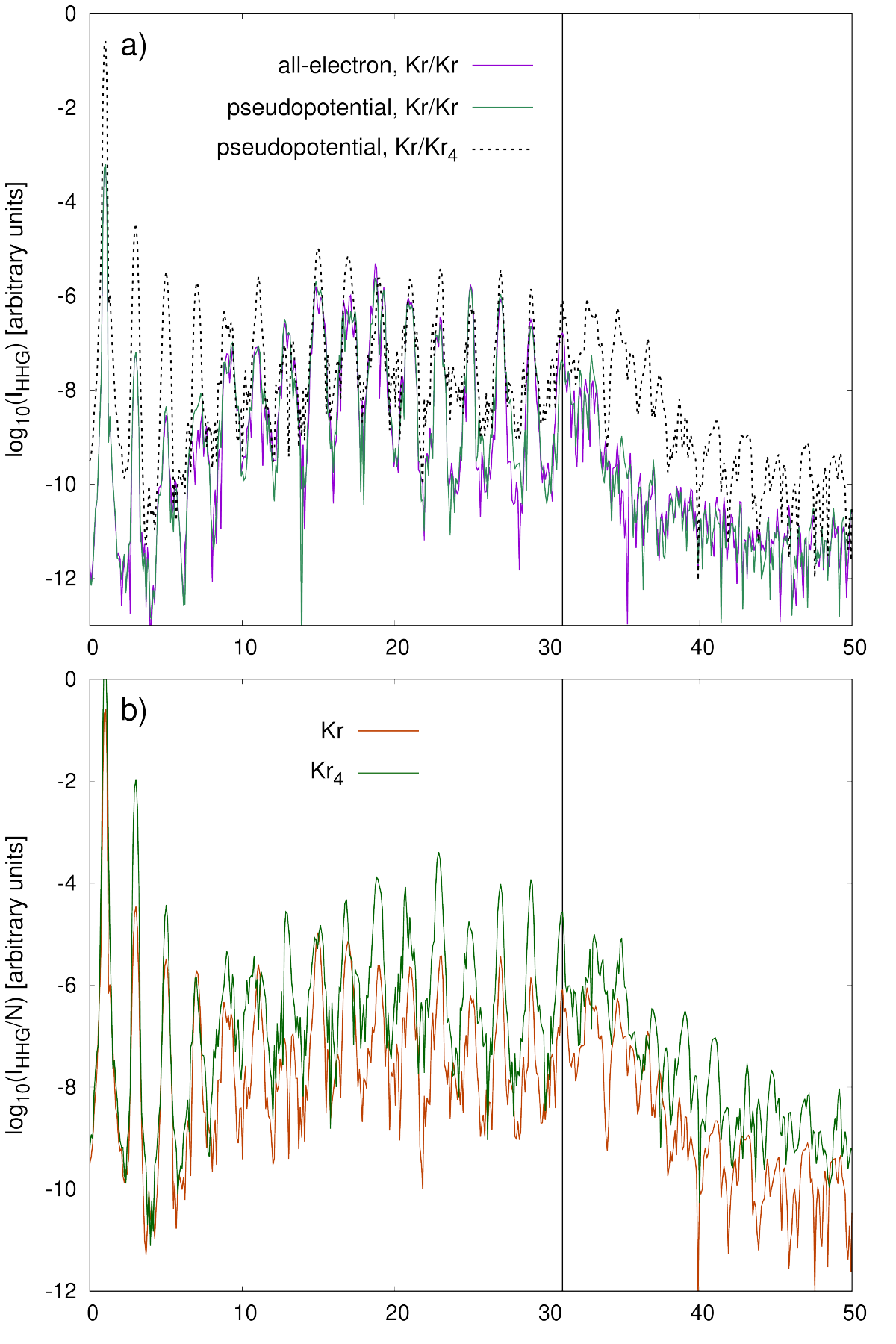}
\caption{a) HHG spectra of atomic krypton computed in the aug-cc-pVTZ+ARO all-electron basis set of a single Kr atom, the aug-cc-pVTZ-PP+ARO basis set of a single Kr atom  coupled to the ECP10MDF pseudopotential, and the aug-cc-pVTZ-PP+ARO basis set of the Kr\textsubscript{4} cluster coupled to the ECP10MDF pseudopotential. b) Computed HHG spectra of Kr\textsubscript{4} and of the Kr atom in the basis of Kr\textsubscript{4}. The black vertical lines denote the 3SM cutoff position}
\label{fig:kr_spectra}
\end{figure}

Since the pseudopotentials used in quantum chemistry have not yet been applied for the purpose of HHG modeling, we first run a preliminary calculations for a single Kr atom in the aug-cc-pVTZ-PP basis set with the ECP10MDF pseudopotential, and in the all-electron aug-cc-pVTZ basis set (both augmented with the ARO functions), at the above laser conditions.
In this way we check whether the inclusion of the pseudopotential has any effects on the obtained HHG spectra.
However, as seen on Fig. \ref{fig:kr_spectra}a, the spectra obtained in both calculations are nearly identical, showing that the excitations from first lowest orbitals, as well as the relativistic effects included implicitly in the pseudopotential, have negligible impact on HHG in krypton.
The reason why we choose krypton and not xenon, even though the latter has even higher propensity for cluster formation, is that the high atomic number of xenon makes its description unreliable at the nonrelativistic level of theory.
Thus, such a comparison would not be possible without introducing relativistic corrections to the all-electron Hamiltonian, which is beyond the scope of this study.
Fig. \ref{fig:kr_spectra}a also shows the effect of adding the ghost atoms on the atomic spectrum of Kr.
While the single-centered basis set performs somewhat better for krypton than it did for helium (Fig. \ref{fig:he_basis}), successfully reproducing all the peaks in the plateau, it still significantly underestimates the HHG intensity, especially in the perturbative region and near the cutoff.

\begin{figure}
\centering
\includegraphics[width=0.99\linewidth]{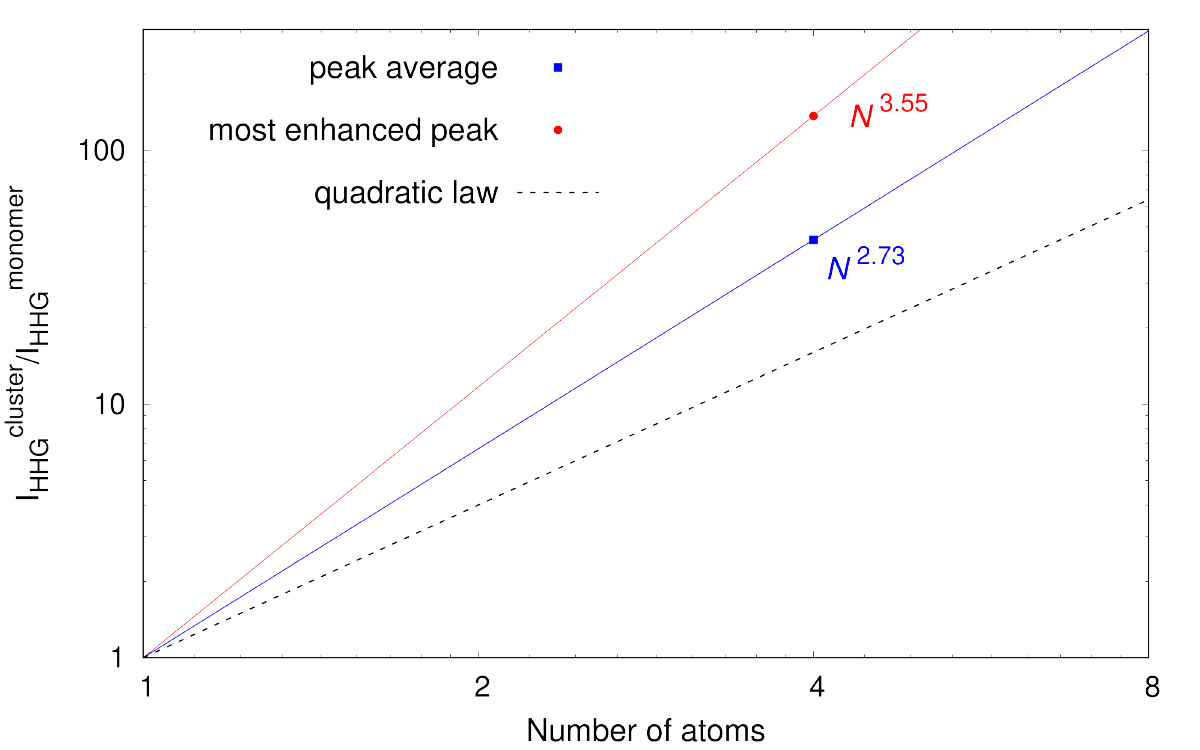}
\caption{Dependence of the HHG intensity of the krypton cluster on the number of atoms, calculated from the average plateau peak intensity and from the intensity of the plateau harmonic that exhibits strongest HHG enhancement}
\label{fig:kr_enhancement}
\end{figure}

The comparison between the spectra of the Kr atom and Kr\textsubscript{4} cluster, both calculated in the basis set of Kr\textsubscript{4}, is presented on Fig. \ref{fig:kr_spectra}b.
Once again it can be noticed that the cutoff position is practically the same in both spectra, and it is in good agreement with the 3SM prediction.
This confirms that the ``atom-to-itself" recombination mechanism is predominant also in clusters of heavier noble gases.
The overall HHG yield of Kr\textsubscript{4} is, as expected, much stronger than of the Kr atom, although the differences in intensities of individual peaks are less uniform than in case of helium clusters.
For two harmonic orders within the plateau region, 11th and 15th, the intensity of the atomic HHG is nearly the same as the intensity of the cluster HHG.
Like in the case of helium clusters, we believe it to be a consequence of the overestimation of ionization by the finite lifetime model within the CIS approximation.
Due to their discrete character, Gaussian basis sets cannot accurately represent a true electronic continuum, but rather approximate it with a series of resonance states.
Consequently, it is possible that certain ionization channels are described with higher accuracy than others and have a more pronounced effect on the ionization loss.
This effect is probably more noticeable in krypton than it was in helium due to its lower $I_p$.
Nevertheless, the analysis of the dependence between the HHG intensity and the number of atoms confirms the presence of the HHG enhancement, with the growth rates of $N^{2.73}$ for the average harmonic intensity and $N^{3.55}$ for the most strongly enhanced peak (Fig. \ref{fig:kr_enhancement}).
Although these values should be regarded as fairly rough estimates, given that we base our predictions on only one cluster size, they are in good agreement with the experimental data.
For comparison, Park \textit{et al.} \cite{park2014} measured a dependence $N^{3.8\pm-0.4}$ for small argon clusers (below 700 atoms) and predicted a similar rate also for neon, krypton and xenon.
Tao \textit{et al.} \cite{tao2017} reported an approximately 100-fold enhancement of the optical response per atom in argon clusters of $N \approx 250$ atoms, with respect to the response of an atom in the gas phase.
When translated to the HHG intensity, this corresponds to an enhancement rate of about $N^{3.7}$ for this cluster size.
Earlier, Tisch \textit{et al.} \cite{tisch1997} showed that xenon clusters exhibit a cubic scaling of the harmonic yield with backing pressure, instead of a quadratic one observed for the monoatomic gas.
Our enhancement rates may be somewhat understated due to the above-mentioned overestimation of ionization.
However, we believe that this error is partially compensated by not including another factors that decrease the experimentally observed HHG intensity, such as the lack of perfect phase matching, partial incoherence of the optical response, or the reabsorption of the emitted radiation.

What is also important, our calculations predict that the HHG yield increase is much stronger in krypton than in helium.
This further supports the proposed polarization mechanism of HHG enhancement, since krypton's higher polarizability and greater number of electrons can be expected to facilitate the shielding of the residual ion and the dissipation of positive charge over surrounding atoms.
All the observations made for the ground states of the He clusters remain valid also in case of Kr\textsubscript{4}.
They include the lack of initial state delocalization (judged by negligible mixing between basis functions centered at different atoms), the additivity of dipole polarizabilities of individual atoms within the clusters, and small differences in the MO energies between a single atom and a cluster.

\section{Conclusion}

To conclude, we reported \textit{ab initio} calculations of the HHG spectra of atoms and small few-atomic clusters of helium and krypton, with the particular focus on the phenomenon of HHG enhancement.
The employed RT-TDCIS method, coupled to Gaussian basis sets designed to describe highly excited and continuum states, and combined with the ghost averaging approach for obtaining the atomic spectra, proved to be able to qualitatively and quantitatively reproduce the ratio between HHG intensities of noble gas atoms and clusters.
What is worth emphasising is the fact that we are able to simulate HHG in both kinds of species using a method that accounts only for single exctitations and single ionization.
We refrain from claiming that multiple excitations play no role in cluster HHG, but our results strongly suggest that it can be approximately described by a mechanism similar to the molecular counterpart of the three-step model, with only one electron per cluster undergoing tunneling ionization.
The calculated growth rates of the harmonic intensity with the number of atoms are between $N^{2.13}$ and $N^{2.76}$ for He clusters, and between $N^{2.73}$ and $N^{3.55}$ for the Kr clusters, with the latter being in good agreement with experimental data for clusters of heavy noble gases.
In the calculations of krypton we also showed that quantum chemical pseudopotentials can be successfully applied to model HHG.

Based on the obtained results, we propose a mechanism explaining the anomalous increase of the emitted harmonic radiation with the number of atoms in small noble gas clusters.
It can be summarized as follows:\\
(i) The initial state of the cluster exhibits little to no delocalization and is approximately the product of ground states of isolated atoms, as shows the analysis of the obtained ground state wavefunctions.\\
(ii) The tunneling ionization and recombination in the cluster occurs according to the ``atom-to-itself" scenario, as can be evidenced by negligible differences between the cutoff positions of atomic and cluster HHG spectra.\\
(iii) Despite the above two points, the cluster acts as a unitary HHG emitter, with only one atom in the cluster being ionized at a time. This explains why it is possible to quantitatively reproduce cluster HHG using the CIS \textit{Ansatz}. In our calculations, due to the high symmetry of examined clusters, the detached electron is in a coherent superposition of states describing ionization from individual atoms.\\
(iv) The additivity of the atomic polarizabilities within the cluster allows for a proportionally larger electronic wavepacket to undergo tunneling ionization, which is responsible for the quadratic part of the relationship between $I_{\mathrm{HHG}}$ and $N$.\\
(v) After the tunneling ionization, the residual ion's charge polarizes the surrounding atoms, increasing the electron density around its own nucleus. The recolliding electron interacts with more bound electrons than in case of an isolated atom, which leads to emission of additional radiation and a higher than quadratic HHG yield increase rate.\\

The above mechanism is highly consistent with the experimental observations and allows for some reinterpretation of the existing data on cluster HHG.
For instance, it suggests that the increase of the polarizability with the atomic number of noble gases is another factor responsible for the fact that HHG enhancement is easiest to notice in heavier gases, apart from their stronger tendency to form clusters.
It is also able to explain the failings of the SAE-based models in simulating cluster HHG.

A natural continuation of this work would be to examine the transition in the HHG mechanism that occurs when moving from the ensemble of individual atoms to a few-atomic cluster.
One of the ways to accomplish it is by simulating the HHG spectra at different interatomic distances.
However, this is a formidable task for at least two reasons.
First, one must employ a wavefunction \textit{Ansatz} that accounts for multiple excitations in order to describe the system in the limit of non-interacting atoms.
Second, a basis set representation is required that can provide the same accuracy in the description of excited and ionized states regardless of the atomic positions.  
Results obtained using localized Gaussian basis sets, on the other hand, are extremely sensitive to displacing or removing orbital centers.
The most desirable approach would therefore be a real-time time-dependent coupled cluster or full configuration interaction with a discrete variable representation (DVR) or grid-based representation of electronic states.
The work in this direction is currently in progress in our group.

\begin{acknowledgments}
The authors thank Leszek Stolarczyk for fruitful discussions and for reading and commenting on the manuscript.
This work was supported by the Polish National Science Centre (NCN) through Grant No. 2017/25/B/ST4/02698. The calculations have been carried out using resources provided by Wroclaw Centre for Networking and Supercomputing, Grant No. 567.
\end{acknowledgments}


\bibliography{apssamp}

\end{document}